\documentclass[conference]{IEEEtran}
\usepackage{cite} 
\IEEEoverridecommandlockouts

\usepackage{cite}
\usepackage{amsmath,amssymb,amsfonts}
\usepackage{algorithmic}
\usepackage{graphicx}
\usepackage{textcomp}
\usepackage{xcolor}
\def\BibTeX{{\rm B\kern-.05em{\sc i\kern-.025em b}\kern-.08em
    T\kern-.1667em\lower.7ex\hbox{E}\kern-.125emX}}
\begin{document}

\title{Integrated tunable magnonic devices for beyond 6 GHz signal processing\\
{\footnotesize \textsuperscript{ }}

}

\author{\IEEEauthorblockN{ M. Cocconcelli, A. Angotti, F. Maspero, A. Cattoni, R. Bertacco}
\IEEEauthorblockA{\textit{Dipartimento di Fisica, Politecnico di Milano, Via G. Colombo 81, 20133 Milano (Italy)} \\
\\
maria.cocconcelli@polimi.it}
}
\maketitle

\begin{abstract}

In the search for novel technology platforms supporting the transition towards "beyond 6G" telecommunications, magnonics is emerging as a viable route, primarly due to its intrinsic compatibility with the  UWB - FR3 bands and its easy tunability. In this paper, we present a proof-of-concept device based on CoFeB magnonic waveguide, fully integrated on silicon, which demonstrates all the key features of our integration approach, based on co-integrated hard magnetic micromagnets, magnonic conduits, reconfigurable soft magnetic elements and MEMS.  The bias field enabling operation up to about 12 GHz is provided by SmCo micromagnets embedded in the silicon substrate. Real-time tunability is implemented in two ways. First, current-driven control is achieved via an integrated current line that generates an additional localized magnetic field, enabling continuous tuning of Backward Volume spin-wave propagation. Second, we achieve voltage-controlled tunability by integrating a NiFeMo soft magnetic element onto a piezoelectric MEMS cantilever and flip‑chipping it onto the magnonic device. Upon actuation, the cantilever brings the NiFeMo into proximity with the permanent micromagnets, where it captures the stray field. The resulting voltage-controlled displacement modulates the coupling between the hard and soft magnetic components, effectively reducing the local bias field on the CoFeB waveguide and enabling fine control of Damon–Eshbach spin waves.
\end{abstract}

\section{Introduction}
Wireless communication technologies are central to modern information systems, spanning applications from smartphones and Internet-of-Things (IoT) devices to satellite communications and radar systems. The continuous increase in data traffic and device density is driving the development of communication hardware capable of operating at higher microwave frequencies while simultaneously providing larger bandwidths, improved spectral selectivity, reduced footprint, and dynamic reconfigurability. In this context, the 7–24 GHz frequency range (FR3) has emerged as one of the most promising spectral windows for future 6G wireless communications, combining substantially larger available bandwidths than conventional sub-7 GHz systems with lower propagation losses than millimeter-wave frequencies \cite{Cui2025}. However, implementing compact and tunable microwave components that can efficiently operate throughout this frequency range remains a major technological challenge. Conventional radio-frequency (RF) filtering and signal-processing technologies, including ceramic monoblock filters as well as surface acoustic wave (SAW) and bulk acoustic wave (BAW) devices, face increasing limitations at these frequencies due to rising insertion losses, limited tunability, electromagnetic interference, and growing integration complexity \cite{Delsing2019,Ilderem2020,Hara2010}. In this context, magnonics has emerged as a promising platform for RF signal processing by exploiting spin waves (SWs), i.e., collective excitations of the magnetic order in magnetic materials \cite{Flebus2024,Chumak2015MagnonSpintronics,Mahmoud2020,Barman2021}.

Spin waves naturally operate over a broad frequency range extending from the MHz regime to the sub-THz regime while offering wavelengths significantly shorter than those of electromagnetic waves at the same frequency, enabling strong device miniaturization \cite{Pirro2021,Chumak2022,Wang2020DirectionalCoupler}. Furthermore, spin-wave propagation is governed by a highly tunable dispersion relation that is strongly dependent on the magnetic environment. The dispersion relation can be tailored through the relative orientation between the spin-wave wavevector and the magnetic bias field, enabling flexible control of propagation characteristics such as phase velocity, attenuation, and wavelength \cite{Stancil2009,Kalinikos1986}. Reconfigurability can additionally be achieved through moderate variations of the local magnetic field, allowing dynamic tuning of spin-wave transmission properties \cite{Vanderveken_2020, Greill_2026, Cocconcelli_2026, Hamalainen2018}.
Despite their potential, practical applications of magnonics in RF systems remain limited. To date, commercial implementations have largely been restricted to yttrium iron garnet (YIG)-based microwave oscillators \cite{RohdeSchwarzFSW} and a limited number of nonlinear microwave components \cite{MetamagneticsAutoTune}. Proposed magnonic devices such as tunable RF filters, phase shifters, and spectrum analyzers have not yet reached widespread technological deployment. Two major obstacles hinder practical integration. First, high-performance magnonic devices usually rely on epitaxial YIG films, which exhibit exceptionally low magnetic damping ($2 \times 10^{-4}$) and long spin-wave propagation lengths but remain difficult to integrate with standard silicon microelectronics \cite{Gaur2023} and are not suitable for real applications requiring wide operating temperature ranges. Just to make an example, the automotive field requires stable performances between -40°C and +125°C, but in this range the saturation magnetization diminishes by about 30\% due to the low Curie temperature of YIG (about 560K), leading to an unacceptable shift of the ferromagnetic resonance frequency of about 0.7 (1.8) GHz for in-plane (out-of-plane) magnetized YIG with applied bias field of 250 mT.  Second, coherent spin-wave excitation and tuning generally requires the application of controllable magnetic bias fields, typically generated by bulky permanent magnets or electromagnets that are incompatible with compact and scalable consumer electronic platforms \cite{Haldar2020}. 

In this work, we address these limitations by demonstrating a proof-of-concept fully integrated, self-biased, and reconfigurable magnonic device realized on silicon using planar processes compatible with wafer-scale mass production. As magnonic material we chose CoFeB for its well mastered compatibility with planar processes on silicon and its high Curie temperature in film with thickness of a few tens of nanometers (up to 700 K). Furthermore, its high saturation magnetization (approximately 10 times larger than that of YIG) enables easy access to the "beyond 6G" frequency range with a relatively low bias field, which can be generated by integrated micromagnets. We introduce a novel method for permanent micromagnet integration, suitable to generate fields as high as 50 mT on the CoFeB magnonic waveguide, substantially larger than previously reported \cite{Cocconcelli2024, Cocconcelli_adv}, enabling spin-wave operation without any externally applied magnetic bias field at frequencies up to 12~GHz. 
We then investigate and experimentally benchmark two complementary strategies for dynamic reconfiguration of the local magnetic landscape in the standalone magnonic platform. The first approach relies on an integrated current line that generates an additional Oersted field superimposed on the static stray field produced by the SmCo micromagnets, thereby enabling continuous electrical tuning of the spin-wave response. Although this strategy is comparatively power consuming, it enables compact device geometries, fast modulation (up to the MHz range), and improved scalability. The second approach exploits voltage-controlled magnetic tuning through a NiFeMo soft magnetic element integrated on a piezoelectric MEMS cantilever \cite{MEMS_ST}, which is subsequently coupled to the magnonic device via flip-chip bonding. In this case, tunability is achieved through controlled displacement of the soft magnetic element, enabling substantially lower power consumption at the expense of a larger device footprint and reduced frequency modulation range (100 kHz). By implementing both approaches on the same self-biased magnonic architecture, a direct comparison between compactness, tunability, and power efficiency is established. Overall the proposed platform emerges as a promising technology offering different routes towards the integration of magnonic elements in real RF devices compatible with CMOS electronics.

\section{Standalone spin-wave propagation enabled by integrated micromagnets}

\begin{figure*}
  \includegraphics[width=\textwidth]{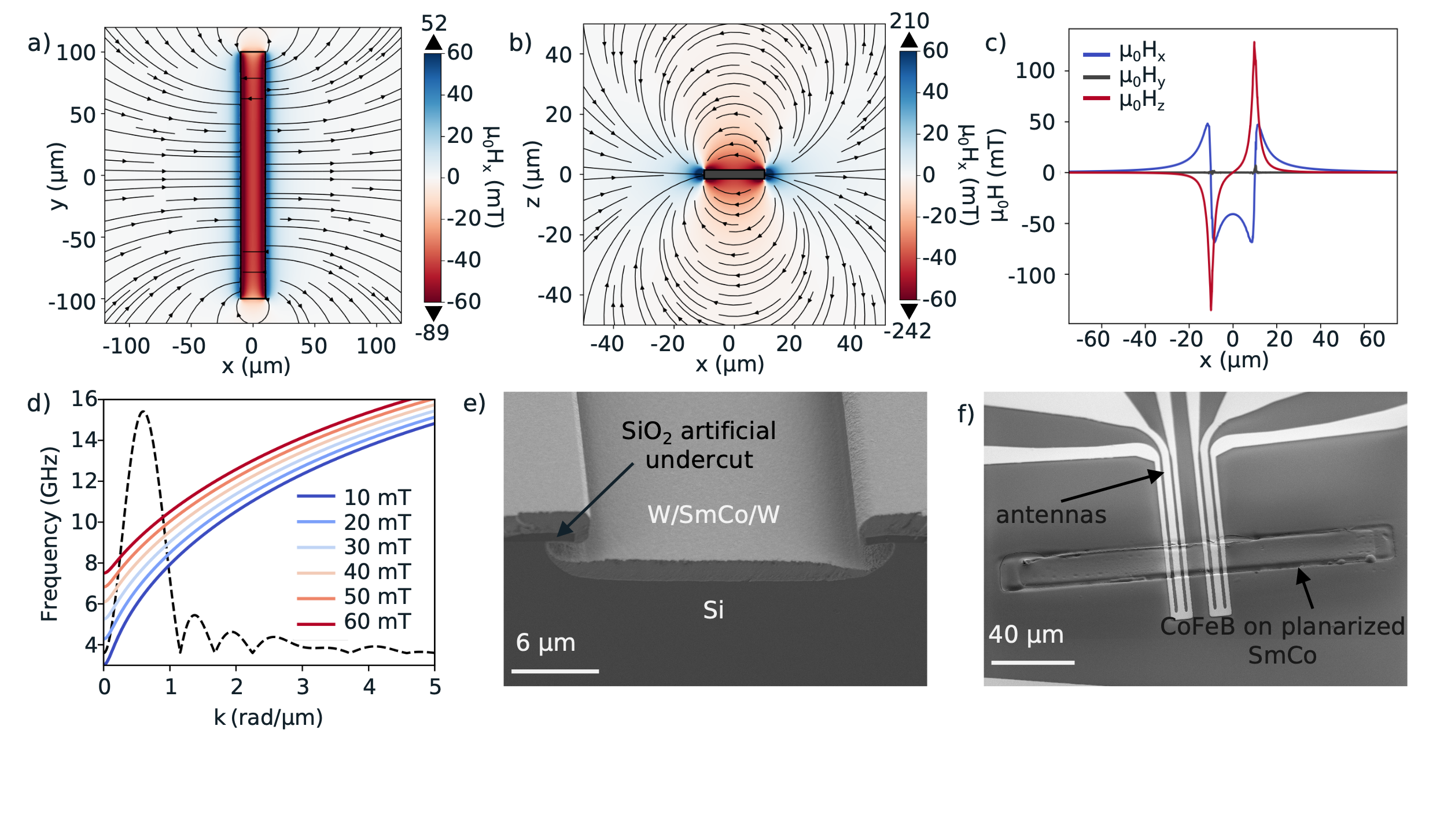}
\caption{a) Simulated $x$-component of the stray field generated by a SmCo micromagnet magnetized along $x$ (color scale), overlaid with stray field lines, shown over the $xy$ plane at a distance of 1~um from the micromagnet surface, corresponding to the position of the magnonic conduit. b) $x$-component of the stray field and stray field lines over the $xz$ plane, showing the cross-section. c) Profile of the stray field components at the height of the CoFeB conduit. d) Damon--Eshbach dispersion relation of the CoFeB conduit calculated for different values of the applied bias field; the black dotted line indicates the antenna efficiency as a function of wavenumber. e) SEM image of the cross-section of the Si pocket in which the micromagnet has been deposited, showing the engineered undercut created by the mesoporous SiO$_2$ sol-gel layer. f) SEM image of the completed device.}
  \label{fig:1}
\end{figure*}
Achieving standalone spin-wave propagation at microwave frequencies requires the generation of a sufficiently strong internal magnetic bias field without relying on externally applied magnets or electromagnets. A possible strategy consists in the integration of permanent micromagnets acting as a source of magnetic field on chip. This imposes simultaneous constraints on both the magnetic material generating the magnetic bias field and the spin-wave conduit supporting propagation. In particular, the magnetic system must provide a stable and localized magnetic field compatible with planar silicon integration, while the spin-wave waveguide must combine low magnetic damping with compatibility with microfabrication technologies.

To generate the internal magnetic bias field, permanent micromagnets based on W(100 nm)/SmCo(3 $\mu$m)/W(100 nm) stacks were employed in this work. SmCo was selected due to its large coercivity and high remanent magnetization, enabling the generation of stable localized stray fields without the need for continuous power consumption. Moreover, SmCo has shown  potential in reaching good magnetic performance, without exceeding the thermal budget of CMOS back-end-of-line (BEOL) processing ($\leq 400^\circ\mathrm{C}$) \cite{cheng_new_2012}. Vibrating sample magnetometry measurements performed on continuous films indicate a saturation magnetization of approximately 0.65 T and a coercive field of approximately 600 mT (see Section S6 of Supplementary Information), ensuring robust magnetic stability against the moderate external fields used during spin-wave characterization.

For the spin-wave conduit, Ta(5 nm)/CoFeB(80 nm)/Ta(5 nm) waveguides were employed. CoFeB combines relatively low magnetic damping, sufficiently large saturation magnetization and compatibility with silicon processing technologies, enabling spin-wave propagation over distances of several tens of micrometers in the GHz frequency range \cite{Yu2012}. Spin waves were excited and detected through coplanar waveguide antennas (CPWs) \cite{Lucassen2019}.

A central aspect of the device design is the co-optimization of the micromagnet geometry, waveguide dimensions, and relative positioning in order to maximize the effective internal magnetic field while preserving efficient spin-wave transduction. In the final architecture, the CoFeB waveguide is positioned directly above the SmCo micromagnet, so that the propagating spin waves experience the localized stray field generated by the permanent magnetic structure. The SmCo micromagnets were designed with lateral dimensions of $20 \times 200~\mu$m$^2$ and magnetized along their short axis. In this configuration, the dominant in-plane component of the stray field is oriented perpendicular to the spin-wave propagation direction (Figure \ref{fig:1}a, b and c), thereby establishing the Damon--Eshbach geometry and maximizing the effective bias field acting within the waveguide.

The dimensions of both the micromagnet and the spin-wave conduit were selected as a tradeoff between operating frequency, field homogeneity, and transmission efficiency. Reducing the lateral dimensions of the micromagnet increases the localized stray field and shifts the spin-wave transmission band toward higher frequencies. However,  lateral confinement increases, thereby modifying the dispersion relation of the confined spin-wave modes together with injection losses (see Section 3 of Supplementary Material).  A width of $20~\mu$m was therefore selected as an optimal compromise between GHz-frequency operation, field uniformity, and acceptable transmission amplitude.

Micromagnetic simulations indicate that the optimized geometry produces an effective internal magnetic field of approximately 50 mT at the waveguide position, as shown in Figure \ref{fig:1}a, b and c, corresponding to a spin-wave transmission band extending approximately from 8 to 12 GHz for the antenna geometry employed in this work, as shown by the analitical calculations of Figure \ref{fig:1}d.

To maximize the effective internal magnetic field while preserving compatibility with planar microfabrication processes, the SmCo micromagnets were embedded within the silicon substrate (Figure \ref{fig:1}e) and the CoFeB waveguide was fabricated directly above them, separated by a mesoporous $1~\mu$m-thick SiO$_2$ sol-gel planarization layer, as shown in Figure \ref{fig:1}f. This configuration minimizes the distance between the permanent magnet and the spin-wave conduit, thereby maximizing the stray field experienced by the propagating spin waves while simultaneously restoring surface flatness and ensuring electrical insulation. More information on the fabrication process can be found in Section 5 of the Supplementary Material.

An important advantage of this architecture is the strong spatial localization of the magnetic bias. Micromagnetic simulations indicate that the stray field generated by the SmCo micromagnets decays almost completely within approximately $50~\mu$m from the magnetic structure (see Figure \ref{fig:1}a, c). As a result, the integrated permanent magnets provide strong local biasing of the spin-wave waveguide without generating large parasitic fields over the surrounding chip area, avoiding the shielding requirements typically associated with macroscopic permanent magnets or external electromagnets.

To experimentally validate standalone spin-wave propagation, vector network analyzer (VNA) measurements were performed on devices both before and after magnetization of the SmCo micromagnets. Frequency sweeps from 4 MHz to 40 GHz were acquired for external magnetic fields $\mu_0 H_{ext}$ ranging from $-100$ mT to $+100$ mT with 1 mT increments. The transmission spectra were processed using reference subtraction and time-domain gating, as detailed in Section 2 of the Supplementary Information.

\begin{figure*}
  \includegraphics[width=\textwidth]{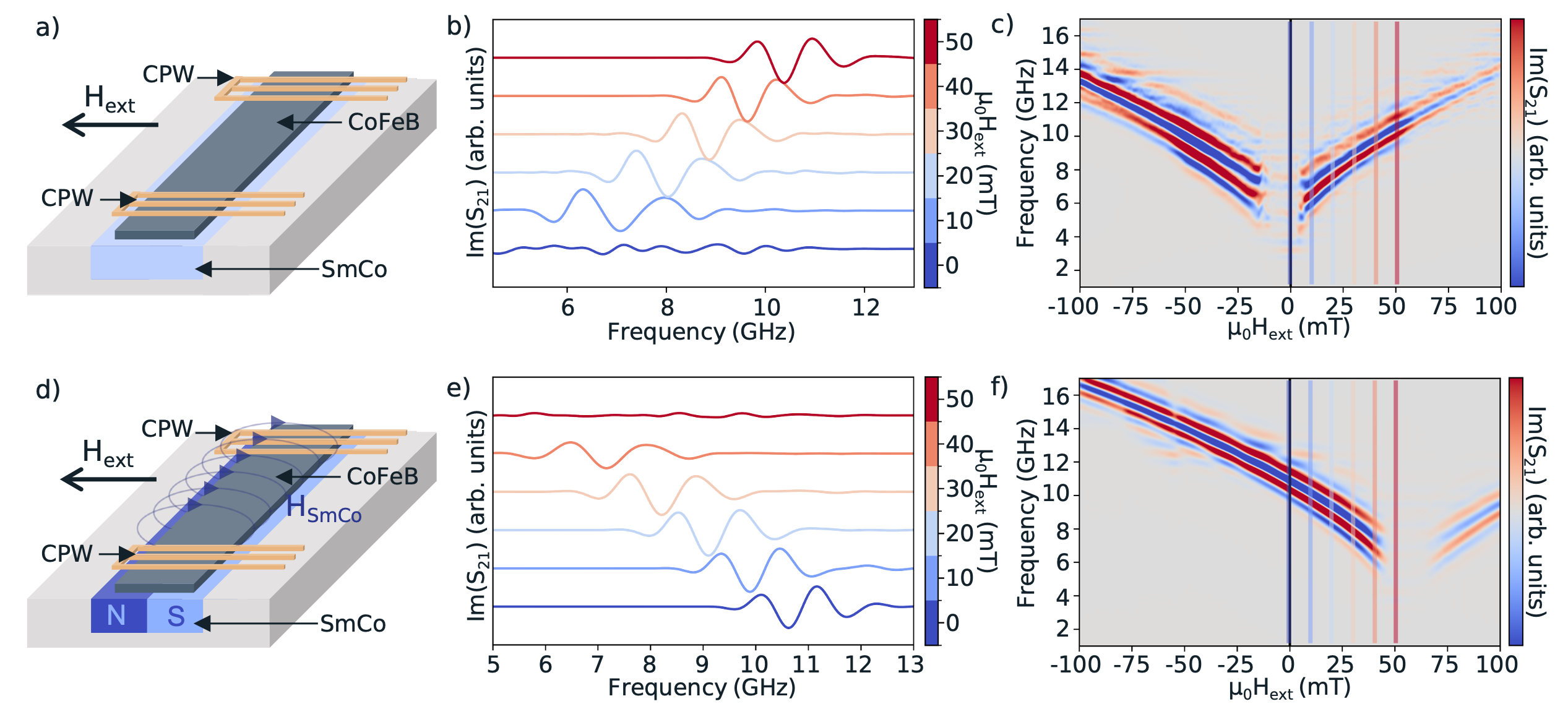}
  \caption{Schematics of the device before (a) and after (d) magnetization (not to scale), illustrating the SmCo micromagnet, the CoFeB magnonic conduit, and the coplanar waveguide (CPW) antennas. Frequency spectra of the reference-subtracted imaginary part of the scattering coefficient $S_{21}$, acquired with a VNA for selected values of the external field $\mu _0H_{ext}$, before (b) and after (e) magnetization. Color maps of the signal as a function of frequency and external magnetic field before (c) and after (f) magnetization.}
  \label{fig:2}
\end{figure*}
The as-fabricated devices, schematized in Figure \ref{fig:2}a, measured prior to magnetization of the SmCo layer, do not sustain spin-wave propagation at zero externally applied magnetic field, as shown in Figure \ref{fig:2}b and c, showing respectively the transmission signal collected at selected values of external field and the frequency versus field map of the spin wave signal. Transmission is observed only for externally applied fields larger than approximately $+10$ mT or smaller than approximately $-20$ mT (Figure \ref{fig:2}c). The asymmetry between positive and negative field values is attributed to a weak remanent magnetization of the SmCo layer. Although the overall magnetization of the unmagnetized micromagnets remains close to zero due to the random orientation of magnetic grains, the annealing process promotes the formation of magnetically harder phases that can produce a small preferential remanence.

Upon magnetization of the SmCo micromagnets using a 2~T external magnetic field (scheme in Figure \ref{fig:2}d), standalone spin-wave propagation is clearly observed in the absence of any externally applied magnetic field, demonstrating fully self-biased operation, as shown in Figure \ref{fig:2}e. The transmission band extends approximately from 9.7 to 12.3 GHz, confirming the presence of a strong internal magnetic bias generated by the permanent micromagnets. When positive external magnetic fields are applied antiparallel to the SmCo stray field, the transmission band progressively shifts toward lower frequencies due to the reduction of the effective internal field within the waveguide, as shown in Figure \ref{fig:2}f.

At external fields close to approximately 50 mT, spin-wave transmission is fully suppressed. This condition corresponds to compensation of the stray field generated by the SmCo micromagnets, resulting in an almost vanishing net magnetic field inside the waveguide and therefore in the loss of a well-defined magnetization state. This measurement provides an experimental estimate of the effective internal bias field generated by the integrated micromagnets, yielding a value of approximately 50 mT, in good agreement with micromagnetic simulations and device design.

For larger positive external magnetic fields, the transmission band shifts back toward higher frequencies while the transmission amplitude decreases. This behavior is consistent with reversal of the magnetization direction in the CoFeB waveguide and with the nonreciprocal nature of Damon--Eshbach spin-wave propagation. Conversely, for negative external magnetic fields, the transmission band continuously shifts toward higher frequencies due to the additive contribution of the external field and the SmCo-generated stray field. In this regime, the external field remains well below the coercive field of the SmCo micromagnets and is therefore insufficient to reverse their magnetization, which consequently remains aligned along its initial direction.

These results demonstrate standalone spin-wave propagation enabled exclusively by the localized stray field generated by integrated permanent micromagnets. The demonstrated architecture enables GHz-frequency magnonic operation without externally applied magnetic bias fields while maintaining strong spatial confinement of the magnetic bias, thereby representing a scalable route toward compact and silicon-compatible magnonic RF systems.

\section{Modulation of spin wave band}
\subsection{Current-driven reconfiguration}
\begin{figure*}
              \includegraphics[width=\textwidth]{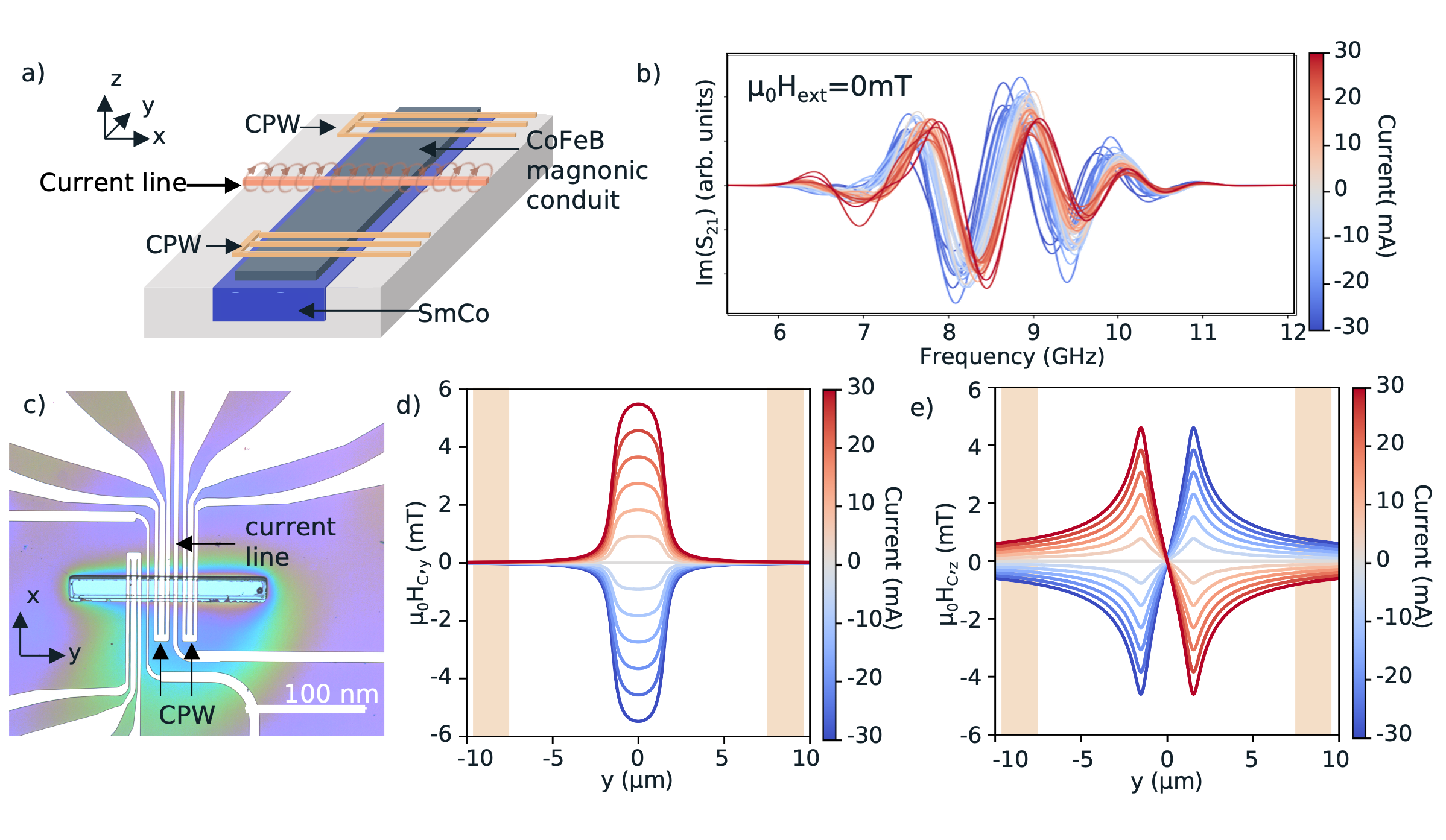}
  \caption{a) Scheme of the current-driven device (not to scale). b) VNA measurements of the tranmission signal of backward-volume spin waves with no magnetic external field applied, applying different currents values. c) Optical image of the device studied, d) and e) y and z components of the calculated magnetic field generated by the current line for different currents applied. The orange bars correspond to the position of the ground of the antennas. }
  \label{fig:3}
\end{figure*}
Dynamic tuning of the standalone magnonic device was first investigated using an integrated current line located between the excitation and detection antennas, as shown schematically in Figure~\ref{fig:3}a and in the optical image in Figure~\ref{fig:3}c. The wire, with a width of $3~\mu$m, generates an Oersted field whose polarity is reversed by inverting the current direction. Micromagnetic simulations of the resulting field distribution are shown in Figures~\ref{fig:3}d and~\ref{fig:3}e, highlighting the in-plane (IP) and out-of-plane (OOP) components responsible for modulating the effective magnetic bias within the spin-wave waveguide. The IP transverse component is highly confined within $5~\mu$m of the conductor, while the OOP peaks are localized at the current line edges and extend partially beneath the antenna regions with a magnitude below 1~mT. Due to the strong IP shape anisotropy of the CoFeB conduit, however, these residual OOP components do not significantly affect spin-wave generation and detection at the antennas. The dominant IP component is instead expected to modify the effective dispersion relation in the region between the antennas, thereby modulating spin-wave transmission.

VNA measurements were performed without any externally applied magnetic field in order to evaluate current-driven modulation of the self-biased spin-wave response. Initial measurements were carried out with the SmCo micromagnet magnetized along its short axis, corresponding to the Damon--Eshbach configuration. In this geometry, the current-induced field mainly produces a tiny local perturbation of the magnetization orientation and no measurable variation in the transmitted signal was observed within the accessible current range.
More pronounced effects were observed when the SmCo micromagnet was magnetized along its long axis, resulting in backward-volume (BV) spin-wave propagation. In this configuration, the stray field generated by the SmCo micromagnet at the waveguide position is lower than in the previously discussed Damon--Eshbach geometry, leading to a spin-wave transmission band extending approximately from 7 to 10.6 GHz (Figure \ref{fig:3}b). The in-plane Oersted field generated by the current line directly adds to or subtracts from the internal bias field produced by the SmCo micromagnet, thereby modifying the effective magnetic field experienced by the propagating spin waves.

Figure~\ref{fig:3}b shows the imaginary part of the measured $S_{21}$ parameter for different applied DC currents. The application of a DC current does not induce a sizeable rigid frequency shift of the transmission band because the excitation and detection regions beneath the antennas are only weakly perturbed by the Oersted field (Fig.~\ref{fig:3}d,e), leaving the excitation conditions essentially unchanged. Instead, the current modifies the pseudo-period of the oscillations in Im($S_{21}$), which decreases as the current (and hence the transverse Oersted field) is swept from negative to positive values.

As shown in previous work \cite{Ciubotaru}, the pseudo-period of the oscillations in Im($S_{21}$) is inversely proportional to the spin-wave group velocity, as it originates from the propagation phase accumulated between the excitation and detection antennas. Therefore, for BV spin waves, whose group velocity increases with the effective bias field, the observed reduction of the pseudo-period can be directly attributed to the increase in the total bias field resulting from the superposition of the static stray field generated by the SmCo magnet and the Oersted field produced by the current-carrying wire.

Overall, these results demonstrate the bidirectional electrical tunability of spin-wave propagation in the standalone magnonic device. The evolution of the frequency-dependent oscillations in Im($S_{21}$) provides direct evidence of the current-induced modification of the spin-wave propagation characteristics. Although the accumulated spin-wave phase cannot be directly extracted from the measured $S_{21}$ phase because of the relatively large insertion losses of the device, it can be estimated from the evolution of the oscillatory features in Im($S_{21}$), as detailed in Section S4 of the Supplementary Material. This analysis provides a convenient quantitative measure of the modulation of spin-wave propagation and, for an antenna separation of 15~$\mu$m, yields a maximum accumulated phase shift of $134^\circ\pm23^\circ$ at 9.5~GHz for a current variation between $-30$ and $30$~mA. Although this tuning scheme requires comparatively large driving currents, it enables compact device geometries and straightforward on-chip integration using conventional microfabrication processes.

\subsection{MEMS-based reconfigurability}
\begin{figure*}
  \includegraphics[width=\textwidth]{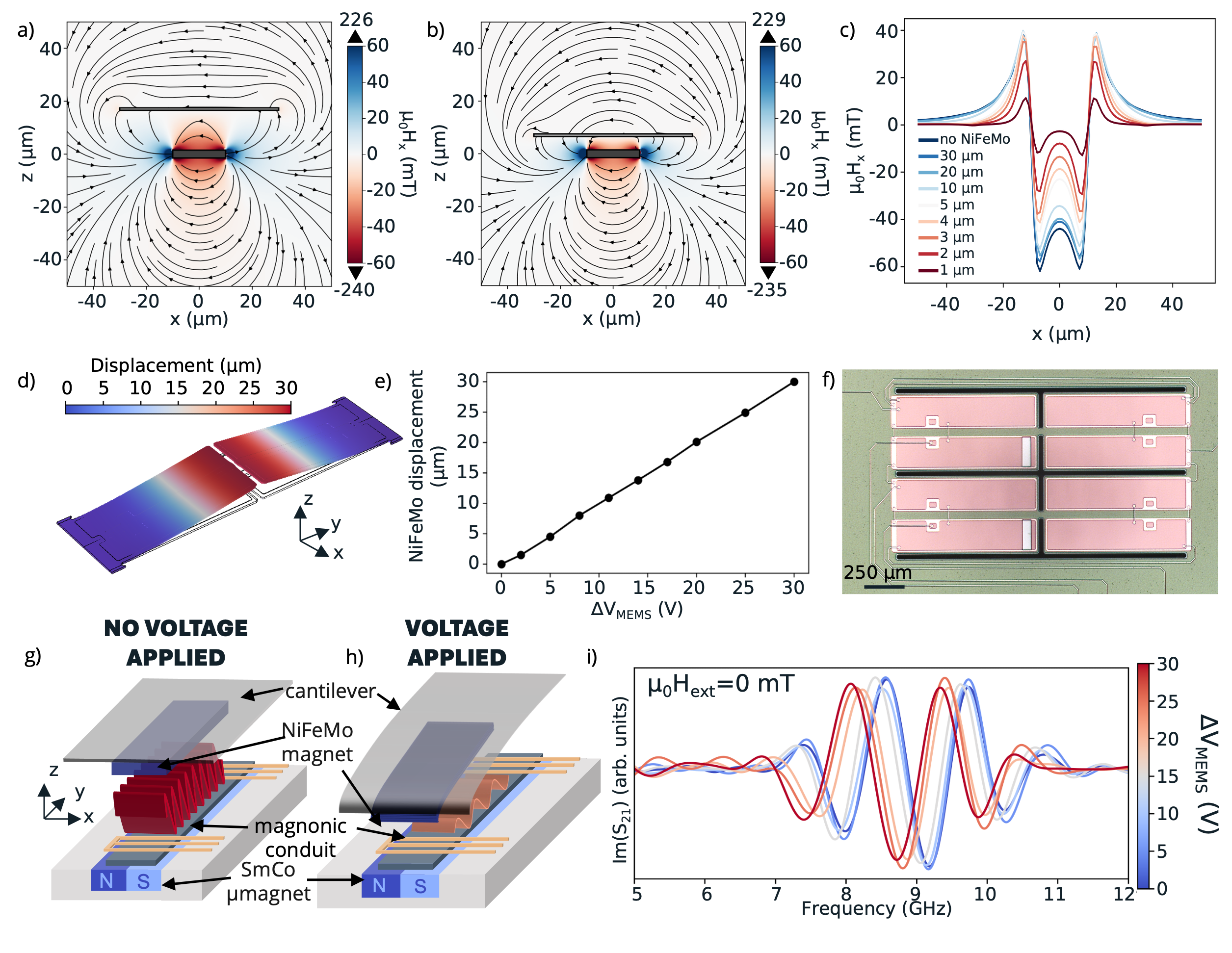}
    \caption{(a,b) Simulated $x$-component of the stray field generated by a SmCo micromagnet magnetized along $x$, shown over the $xz$ cross-sectional plane, with a NiFeMo micromagnet located 15~$\mu$m (a) and 5~$\mu$m (b) away, obtained with COMSOL Multiphysics. (c) $x$-component of the stray field at the CoFeB layer position as a function of the distance between the SmCo and NiFeMo micromagnets. (d) COMSOL Multiphysics simulation of the bending profile of two cantilevers at $\Delta V = 30$~V upon removal of the central supporting bridge. (e) Cantilever tip displacement measured by optical profilometry as a function of applied voltage. (f) Optical image of the two MEMS devices (piezo-pads in pink) after patterning of the NiFeMo micromagnets (vertical white rectangles). (g,h) Schematics of the device at zero applied voltage (g) and under applied voltage (h) (not to scale). (i) Imaginary part of the transmission parameter Im($S_{21}$) for Damon--Eshbach spin waves, acquired in zero external magnetic field for different values of the applied voltage.}
  \label{fig:4}
\end{figure*}
Voltage-controlled tuning was investigated using a hybrid-integrated MEMS actuator incorporating a NiFeMo (Supermalloy) soft magnetic element positioned above the standalone magnonic device. In this approach, tunability is achieved through dynamic reconfiguration of the localized magnetic landscape experienced by the spin-wave waveguide. The soft magnetic element becomes magnetized by the stray field generated by the SmCo permanent micromagnets and partially redirects the magnetic flux, thereby reducing the effective magnetic field acting on the CoFeB waveguide, as illustrated by the magnetic-field simulations in Fig.~\ref{fig:4}a--c. The NiFeMo element is integrated onto a piezoelectric MEMS cantilever whose vertical displacement is controlled through the application of an external voltage. The simulated cantilever deformation and the measured tip displacement are shown in Fig.~\ref{fig:4}d, e, while an optical image of the MEMS device is reported in Fig.~\ref{fig:4}f. The operating principle of the device in the unbiased and actuated configurations is schematically illustrated in Fig.~\ref{fig:4}g,h. As the cantilever approaches the device surface, the magnetic flux capture becomes stronger, progressively reducing the internal magnetic bias field within the waveguide and consequently shifting the spin-wave transmission band toward lower frequencies. This mechanism enables low-power voltage-controlled modulation of spin-wave propagation without the need for externally applied magnetic bias fields.

The MEMS chip was integrated face to face onto the standalone magnonic platform through flip-chip thermal compression bonding (more details in Section S7 of Supplementary Material). VNA measurements performed after assembly confirm that spin-wave propagation remains sustained in the absence of externally applied magnetic fields (see Figure \ref{fig:4}i), demonstrating preservation of the self-biased operation after hybrid integration.

Compared to measurements acquired prior to flip-chip bonding, the spin-wave transmission band is shifted by approximately 2.2~GHz toward lower frequencies (see Figure~S8 of the Supplementary Material). This behavior indicates a reduction of the effective internal magnetic field within the waveguide, which may originate from partial demagnetization of the SmCo micromagnets during thermal compression bonding, from magnetic flux redistribution induced by the NiFeMo element, or from a combination of both effects. Indeed, after flip-chip bonding, the NiFeMo magnet is located only $30~\mu$m from the magnonic chip surface — a distance determined by the bump height — and will approach it further as voltage is applied.

The effect of MEMS actuation on spin-wave propagation was investigated by performing VNA measurements at zero externally applied magnetic field while varying the actuation voltage. Figure~\ref{fig:4}i shows the imaginary part of the measured $S_{21}$ parameter for increasing MEMS voltages. At variance with the case of the current-driven modulation, the perturbation of the SmCo stray field by the NiFeMo soft magnet is quite uniform and involves also the antenna region. By consequence, as the voltage increases the whole spin-wave transmission band progressively shifts toward lower frequencies. This is  consistent with a reduction of the effective internal magnetic field produced by the MEMS-controlled magnetic element.
A maximum frequency shift of approximately 400 MHz was achieved, corresponding to an estimated variation of the effective internal field of about 20 mT. For actuation voltages exceeding approximately 30 V, no additional frequency shift is observed, indicating saturation of the mechanical displacement due to contact between the MEMS structure and the underlying chip. Importantly, spin-wave propagation remains fully observable in this regime. Upon reducing the applied voltage, the transmission band reversibly shifts back toward higher frequencies, demonstrating stable and reversible operation without observable mechanical degradation.

As for the current-driven device, the accumulated spin-wave phase can be estimated from the evolution of the oscillatory features in Im($S_{21}$) (see Section S4 of the Supplementary Material). This analysis yields an estimated phase variation approaching $122^\circ\pm17^\circ$ between 0 and 30~V actuation at 9.5~GHz. These results further demonstrate that MEMS-mediated modulation of the local magnetic field provides effective voltage-controlled tuning of self-biased spin-wave propagation. Although the present device is limited by insertion losses and actuation range, the demonstrated architecture establishes a pathway toward low-power, electrically reconfigurable magnonic phase shifters compatible with hybrid-integrated silicon technologies.

\section{Discussion}

The two reconfiguration strategies investigated in this work rely on the same underlying principle, namely the dynamic modulation of the localized internal magnetic field generated by the integrated SmCo micromagnets. However, the two approaches operate in substantially different regimes in terms of tunability mechanism, device compactness, and power consumption.

The current-driven approach achieves tuning through the local Oersted field generated by an integrated wire positioned close to the spin-wave waveguide. This strategy enables compact device geometries, with an active area on the order of $200 \times 200~\mu$m$^2$, and provides bidirectional field tunability through reversal of the current polarity. A maximum phase shift of about $134^\circ\pm23^\circ$ has been observed when sweeping the current in the wire from -30 to +30 mA.

In contrast, the MEMS-based approach exploits voltage-controlled displacement of a soft magnetic element to modulate the effective magnetic field experienced by the spin-wave waveguide. Although the demonstrated prototype occupies a larger footprint due to the MEMS die size, this approach enables lower-power operation together with frequency shifts approaching 400 MHz and estimated phase variations of about $122^\circ\pm17^\circ$.

The most significant distinction between the two approaches lies in their energy efficiency. In the current-driven device, modulation requires DC currents on the order of 30 mA flowing through a micron-scale Au current line with a resistance of approximately 80 $\Omega$, resulting in Joule dissipation on the order of $I^2R \approx 70$ mW. In contrast, the MEMS-based approach relies on piezoelectric actuation, where static operation is dominated by leakage currents below 1 nA, corresponding to intrinsic power consumptions on the order of 10 nW. Even including the external driving electronics, with power consumption of approximately 150 $\mu$W, the MEMS-based strategy remains orders of magnitude more energy efficient than current-driven modulation.

An additional difference between the two approaches lies in their achievable modulation speed. In the current-driven architecture, the magnetic field is generated directly by the Oersted field of the injected current and can therefore, in principle, be modulated at frequencies extending well beyond the MHz range, primarily limited by the electrical bandwidth of the driving circuitry, parasitic impedances, and thermal constraints. This makes current-driven tuning particularly attractive for high-speed dynamic reconfiguration.

In contrast, the modulation bandwidth of the MEMS-based approach is intrinsically limited by the mechanical response of the piezoelectric actuator. In the present devices, mechanical resonances are observed up to frequencies on the order of several tens of kHz, with actuation efficiency progressively decreasing beyond the mechanical resonance regime. Consequently, MEMS-mediated tuning is better suited for low-power quasi-static or slowly reconfigurable magnonic systems rather than ultrafast modulation applications.

These results highlight a multidimensional trade-off between compactness, energy efficiency, and modulation speed in reconfigurable self-biased magnonic systems. Current-driven tuning provides compact, fast, and scalable on-chip integration at the expense of larger electrical power dissipation and Joule heating. Conversely, MEMS-mediated tuning enables extremely low-power operation and larger magnetic-field modulation amplitudes, although presently with increased footprint, slower modulation speed due to the mechanical response of the actuator, and greater fabrication complexity associated with hybrid integration.

Several aspects of the proof-of-concept device presented here remain to be optimized. In particular, RF insertion losses should be reduced to below 10~dB to meet the requirements of practical RF systems. This can be achieved by increasing the CoFeB thickness (see Section~S3 of the Supplementary Material) and by optimizing the antenna design, which could enhance spin-wave excitation and detection efficiency~\cite{Erdelyi2025}, increase the accumulated propagation phase, and improve the signal-to-noise ratio. For the MEMS-based devices, dedicated miniaturized actuators could be used to substantially reduce the overall footprint while preserving low-power operation. Likewise, optimization of the current-line geometry would improve the efficiency of Oersted-field generation while reducing Joule losses.

Despite these opportunities for further optimization, the present work establishes the key building blocks of a self-biased, reconfigurable magnonic platform operating in the microwave regime.
this work clearly demonstrate the integration of all relevant features of a novel technology platform we are developing to offer a viable route for "beyond 6G" applications: (i) wide-band operation up to 12 GHz, (ii) full integration and standalone operation without need of external source of magnetic fields, (iii) low-power tunability using micromagnets on MEMS, (iv) fast tunability using current lines, (v) wide operating temperature range thanks to the high Curie temperature of SmCo and CoFeB, (vi) excellent robustness against external perturbation thanks to the high coercive field of SmCo integrated permanent magnets.

\section{Conclusion}

In this work, we demonstrated compact, integrated, and tunable RF magnonic devices realized with planar technology on silicon, either monolithically or via SoC integration. The localized stray field generated by SmCo permanent micromagnets embedded in the silicon substrate provides effective magnetic biasing of CoFeB spin-wave waveguides, enabling GHz-frequency operation without external sources of magnetic fields. Standalone spin-wave propagation was experimentally demonstrated in the 9.7--12.3~GHz range, corresponding to an effective internal bias field of approximately 50~mT.

Building on this platform, two complementary approaches for dynamic magnetic reconfiguration were investigated. Current-driven tuning using integrated current lines enables bidirectional modulation of the spin-wave phase up to about $134^\circ \pm 23^\circ$, while MEMS-mediated tuning based on displacement of a soft magnetic element enabled voltage-controlled frequency shifts approaching 400~MHz together with estimated phase variations of $122^\circ \pm 17^\circ$.

The two approaches exhibit complementary advantages. Current-driven modulation enables compact, scalable, and fast integration, with modulation speeds potentially extending far beyond the MHz range, whereas MEMS-mediated magnetic tuning provides substantially lower power consumption and larger magnetic-field modulation amplitudes, although with slower response times limited by the mechanical dynamics of the actuator. Importantly, both strategies preserve the core advantage of the proposed architecture, namely operation without externally applied magnetic bias fields through the use of highly localized integrated permanent micromagnets.

These results establish localized permanent-magnet biasing combined with electrical magnetic reconfiguration as a viable route toward compact, reconfigurable, and silicon-compatible magnonic RF devices for future microwave signal-processing applications applicable for 6G.

\section{Methods}
\textbf{Device Fabrication} Device fabrication was carried out on Si substrates in which trenches were defined by isotropic reactive ion etching to a depth of $4~\mu$m using a bilayer resist stack consisting of an inorganic SiO$_2$-based resist and an organic resist. Subsequently, a W(100 nm)/SmCo(3 $\mu$m)/W(100 nm) multilayer stack was deposited by magnetron sputtering. Patterning was achieved by dry lift-off, while crystallization of the SmCo layer was obtained through a post-deposition annealing step at $650,^{\circ}$C for 30 min. Residual SiO$_2$ debris was removed by immersion in a 1\% HF solution. Additional details regarding the fabrication and integration of the permanent magnetic material are provided in Section S5 of Supplementary Material.

To reduce surface topography and restore surface flatness, a planarization step was performed using a mesoporous SiO$_2$ sol-gel layer with a thickness of approximately 1 $\mu$m, resulting in a surface roughness below 1 nm RMS.

The magnonic waveguides were subsequently fabricated by magnetron sputtering of a Ta(5 nm)/CoFeB(80 nm)/Ta(5 nm) trilayer stack, followed by the deposition of a 50 nm-thick SiO$_2$ insulating layer. RF transducers were then defined by photolithography, electron-beam evaporation, and lift-off of a Ti(10 nm)/Au(100 nm) metallization stack. For the current-driven devices, the integrated current lines were patterned during the same lithographic step. For the MEMS-integrated devices, the metallization layer instead included the electrical pads required for subsequent flip-chip bonding.

For MEMS integration, piezoelectric cantilevers provided by STMicroelectronics were employed \cite{MEMS_ST}. The soft magnetic elements were fabricated directly on the suspended cantilever structures by sputtering a Cr/NiFeMo multilayer stack with a total thickness of the magnetic layer of $1~\mu$m, followed by lift-off patterning. Gold bumps were subsequently deposited on the MEMS contact pads to enable electrical connection to the top and bottom electrodes of the piezoelectric actuator. Finally, the MEMS chip was integrated onto the magnonic platform by flip-chip thermal compression bonding, simultaneously establishing mechanical and electrical connections while maintaining a nominal separation of approximately $30~\mu$m between the MEMS structure and the magnonic device.

\textbf{Simulations and Modelling} Micromagnetic simulations and theoretical modelling were used to evaluate the magnetic stray field generated by the SmCo micromagnets and its interaction with the soft magnetic NiFeMo element used for MEMS-based tuning. The simulations were carried out using the magnetostatic module of COMSOL Multiphysics 6.4.

\textbf{MEMS Characterization} The out-of-plane displacement of the MEMS cantilevers after flip-chip integration was characterized using an optical profilometer (Filmetrics Profilm3D) prior to flip chip bonding. Measurements were performed as a function of the applied actuation voltage to establish the displacement-voltage characteristics of the device. Further information on the MEMS characterization can be found in Section S1 of the Supplementary Material. 

\textbf{Broadband Spin-Wave Spectroscopy} Spin-wave transmission measurements were performed using a RF probe station equipped with a quadrupolar vector electromagnet capable of generating in-plane magnetic fields up to 200 mT, together with a four-port vector network analyzer (R\&S ZNA43) operating up to 43 GHz. Measurements were conducted using frequency-domain acquisition. The excitation frequency was swept from 4 MHz to 40 GHz using 10,000 equally spaced frequency points, resulting in a harmonic frequency grid with a spacing of $\delta f = 4$ MHz. The uniform frequency sampling enabled subsequent time-domain transformations and time-gating procedures. During all measurements, an RF power of 0 dBm was applied and the complete scattering matrix was acquired. More information on the methodology followed can be found in Section S2 of the Supplementary Material. 

Current-driven and voltage-driven tuning experiments were performed using a Keithley 2450 source-measure unit to supply the DC current or actuation voltage, respectively.

For the MEMS-based tuning experiments, a dedicated measurement sequence was adopted to minimize the influence of piezoelectric hysteresis. Prior to data acquisition, the piezoelectric actuator was polarized by applying a voltage larger than the maximum value used during the measurement sequence. The voltage was then reduced to the minimum value of the investigated range. At each voltage step, the complete scattering matrix was acquired while sweeping the magnetic field across the selected range. Only after completion of the entire magnetic-field sweep was the actuation voltage increased to the next value. This procedure ensured that all measurements acquired at a given voltage corresponded to a well-defined and reproducible mechanical configuration of the MEMS actuator, thereby minimizing artifacts associated with hysteretic mechanical behavior.

\medskip
\textbf{Supporting Information} \par 
Supporting Information is available from the authors.

\medskip
\textbf{Acknowledgements} \par 
The authors acknowledge funds from the European Union via the Horizon Europe project ``MandMEMS'', grant 101070536, and from NextGenerationEU, PNRR MUR -- M4C2 -- Investimento 3.1, project IR\_0000015 -- ``Nano Foundries and Fine Analysis -- Digital Infrastructure (NFFA--DI)'', CUP B53C22004310006. 
We thank P.~Pirro for general discussions on the applications of our platform; L.~Castoldi and S.~Zerbini from STMicroelectronics for providing MEMS devices; B.~Heinz from RPTU Kaiserslautern-Landau for his support with focused ion beam etching; and F.~Mancarella and F.~Tamarri from CNR Bologna for their support with preliminary RIE process development. 
This work has been partially performed at Polifab, the micro- and nanofabrication facility of Politecnico di Milano.

\bibliographystyle{IEEEtran}
\bibliography{bibliography}
\end{document}